\documentclass[aps,amsmath,prd,floatfix
]{revtex4}

\usepackage{graphicx}

\newcommand{\gras}[1]{\mbox{\boldmath $#1$}}
\newcommand{\be}{\begin{equation}}
\newcommand{\ee}{\end{equation}}
\newcommand{\ba}{\begin{eqnarray}}
\newcommand{\ea}{\end{eqnarray}}
\newcommand{\Mc}{{\cal M}}
\newcommand{\Ms}{M_{\odot}}
\newcommand{\m}{\langle}
\newcommand{\M}{\rangle}

\newcommand{\bml}{\begin{mathletters}}
\newcommand{\eml}{\end{mathletters}}

\def\ltsima{$\; \buildrel < \over \sim \;$}
\def\simlt{\lower.5ex\hbox{\ltsima}}
\def\gtsima{$\; \buildrel > \over \sim \;$}
\def\simgt{\lower.5ex\hbox{\gtsima}}
\begin{document} 

\title{The effect of the LISA response function on observations of
monochromatic sources}

\author{Alberto Vecchio and Elizabeth~D.~L.~Wickham}
\affiliation{School of Physics and Astronomy, 
University of Birmingham, Edgbaston, Birmingham B15 2TT, UK}

\date{\today}
\begin{abstract}
The Laser Interferometer Space Antenna (LISA) is expected to provide the
largest observational sample of
binary systems of faint sub-solar mass compact objects, in particular 
white-dwarfs, whose radiation is monochromatic over most of the LISA
observational window. Current astrophysical estimates suggest that the 
instrument will be able to resolve $\sim 10^4$ such systems, with a large fraction
of them at frequencies $\simgt 3$ mHz, where the wavelength of gravitational waves 
becomes comparable to or shorter than the LISA arm-length. This affects the structure of
the so-called LISA transfer function which cannot be treated as constant in this frequency 
range: it introduces characteristic phase and amplitude
modulations that depend on the source location in the sky and the emission frequency. 
Here we investigate the effect of the LISA transfer function on detection and parameter
estimation for monochromatic sources. For signal detection we show that filters constructed
by approximating the transfer function as a constant (long wavelength approximation)
introduce a negligible loss of
signal-to-noise ratio -- the fitting factor always exceeds 0.97 -- for $f \le 10$ mHz, 
therefore in a frequency range where one would actually expect the approximation to fail. 
For parameter estimation, we conclude that in the range $3 \,{\rm mHz} \simlt f \simlt 30$ mHz
the errors associated with parameter measurements differ from $\simeq 5\%$ up to a factor $\sim 10$ (depending
on the actual source parameters and emission frequency) with respect to those computed
using the long wavelength approximation.
\end{abstract}

\maketitle

\section{Introduction}
\label{sec:intro}

The Laser Interferometer Space Antenna (LISA) is a space borne laser interferometer 
of arm-length 5 million km for the observation of gravitational waves (GWs) in the 
frequency window $10^{-4}$ Hz - 0.1 Hz~\cite{lisa_ppa}. Amongst the great 
variety of sources that the instrument will be monitoring~\cite{CT02}, binary 
systems of faint sub-solar mass compact objects, in particular 
white-dwarfs, will be considerably abundant~\cite{HBW90,NYPZ04,BH97,NYP01}. In fact LISA 
is expected to provide the largest observational
sample of these faint stars; current estimates suggest that $\sim 10^4$ systems 
will be resolvable~\cite{NYP01}, including about a dozen of already known galactic binary
systems (the so-called "verification sources") easily detectable during the first 
few weeks of operation. Considerable attention has been 
recently devoted to (sub-) solar mass binary systems, as they will allow us to 
investigate the evolutionary history of degenerate stars, their
formation rate, distribution in the galaxy and 
mass transfer~\cite{CT02,Cutler98,TS02,Seto02MNRAS}. Key issues in
preparation of LISA are the design of appropriate data analysis 
schemes to extract effectively and efficiently signals from noise and 
the investigation of the astronomical 
information that can be gathered from the LISA data set, possibly followed up by 
observations with optical telescopes~\cite{CFS04,CS04}.

Sub-solar mass binary systems are expected to produce a moderate-to-large
signal-to-noise ratio in the LISA data set
and their radiation is monochromatic over most of the instrument sensitivity window, 
{\em i.e.} the intrinsic
frequency drift during the observation time $T\approx 1$ yr is smaller than
the frequency resolution bin of width $\Delta f = 1/T$ (this of course ignores the 
spreading of power in adjacent frequency bins induced by the LISA orbital motion 
around the Sun and the change
of orientation of the detector with respect a putative source). 
In the context of data analysis, both signal detection and parameter estimation, 
the response of LISA to gravitational waves
plays a vital role because it introduces features, such as 
amplitude and phase modulations, that need to be properly accounted for during signal
processing.
For LISA, which observes gravitational radiation from binary systems over a frequency
band where the signal wavelength $\lambda$ can be either longer or shorter than
the interferometer arm length, $L = 5\times 10^6\,{\rm km}$, it is well
known that the detector response changes dramatically at 
$f \sim f_{\ast} \equiv 1/(2\,\pi L)\simeq 9.6$ mHz, which corresponds to
the inverse of the round-trip light-time along the LISA arms~\cite{CR03}. If
$f \ll f_{\ast}$ (long-wavelength regime), the instrument transfer function can 
be effectively regarded as constant while
for $f \simgt f_{\ast}$ the transfer function depends on the
source's instantaneous emission frequency and location in the sky, which in turn
introduces time and
frequency dependent amplitude and phase modulations at the
detector output. The LISA transfer
function, then, starts to substantially change the structure of the LISA Michelson 
observable right at the heart of the instrument's sensitivity window $f \approx 3$ 
mHz~\cite{RCP03}, and yet the implications for data analysis have been so far 
largely ignored. In the context of signal detection, no study has been 
devoted to the identification of the frequency range over which it is indeed
safe (with respect to losses of signal-to-noise ratio) 
to approximate the transfer function as a constant in generating signal
templates (which would clearly reduce 
the complexity of the problem at hand). As
far as parameter estimation is concerned, a number of studies have been carried out so far
for monochromatic sources aimed at computing lower bounds to the
the errors associated with parameter measurements~\cite{Cutler98,Seto02MNRAS,TS02,CV98}. 
However, in all of them the LISA transfer function has been treated as a constant.
Such a simplifying assumption is likely to affect the evaluation of the actual
accuracy with which source parameters can
be measured, and has already been pointed out in the context of observations
of massive black hole binary systems~\cite{Seto02PRD}. 

The goal of this paper
is to investigate the effect of the LISA transfer function on signal detection
and parameter estimation for monochromatic sources. In particular
(i) we identify the frequency window over which signal templates can be
safely constructed by approximating the transfer function as a constant
(i.e. the signal-to-noise ratio is very marginally affected) and
(ii) explore the implication of the frequency dependent transfer function
on the errors of parameter measurements.
For signal detection, the main outcome of our work is that one can safely
work in the long-wavelength approximation well into the regime where
one would actually expect the approximation to fail: effectively the use
of templates computed using the long-wavelength approximation does not introduce 
any significant loss of signal-to-noise ratio up to $f \approx 10$ mHz. 
For parameter estimation we conclude that at $f \approx 5$ mHz the errors are already 
starting to depart, by $\approx 5\% - 10\%$, from those computed using
the long wavelength approximation. Such discrepancy becomes more pronounced, 
on average, as the
signal emission frequency increases, and in the frequency range $10\,{\rm mHz} \le f 
\le 30$ mHz the errors are considerably smaller (up to a factor $\sim 10$ for the
3-armed LISA) than the ones previously 
reported in the literature.
 
The paper is organised as follows: in Section~\ref{sec:output} we review 
the signal detected at the output of the LISA Michelson interferometer 
in the so-called long-wavelength approximation (constant transfer function) and rigid
adiabatic approximation (frequency dependent transfer function);
Sections~\ref{sec:detect} and ~\ref{sec:estimate} contain the key results of the
paper: in Section ~\ref{sec:detect} we estimate the degradation of signal
to noise ratio (as a function of frequency) introduced by searching for signals  
using templates that approximate the transfer function as a constant; 
in Section~\ref{sec:estimate}
we show the effects of the LISA transfer function on parameter estimation by
computing the inverse of the Fisher information matrix; in both cases we 
perform extensive Monte Carlo simulations in order to sample a wide parameter space; 
Section~\ref{sec:concl} contains our conclusions and pointers to future work.

\section{The signal measured at the LISA output}
\label{sec:output}

LISA consists of a constellation of three drag-free spacecraft 
placed at the vertices of an ideal equilateral triangle 
with sides $L \simeq 5\times 10^6\,{\rm km}$ to form
a three-arm interferometer, with a $60^\circ$ angle between two 
adjacent laser beams. The barycentre of 
the instrument follows an almost circular heliocentric orbit (the eccentricity
is $< 0.01$), $20^\circ$ behind the Earth; the detector plane is tilted by 
$60^\circ$ with respect to
the Ecliptic and the instrument counter-rotates around the normal to the 
detector plane with the same period $1\,{\rm yr}$ (we refer the reader 
to~\cite{lisa_ppa,Cutler98,RCP03} and references therein for more details). Due to the
finite arm length, the round-trip light-travel
time between two vertices 
of the constellation is finite and this introduces a characteristic frequency
defined as:
\be
f_{\ast} \equiv \frac{1}{2\,\pi{\rm L}}\simeq 9.6\times 10^{-3}\,{\rm Hz}\,.
\label{fast}
\ee

The LISA
``Michelson output'' is synthesised in software using a technique
known as Time Delay Interferometry (TDI) by combining the Doppler readouts registered
at several points in the LISA constellation~\cite{TAE01}. The eccentricity
of the orbit, the motion of the spacecraft and the finite arm length
of LISA affect, in a non trivial way, the detector output generated
by metric perturbations induced by impinging GWs (cf Eq. (11)-(39) of~\cite{RCP03}). 
However, two fairly
simple approximations to the {\em exact} LISA output have been derived:
(i) the {\em long wavelength approximation} and (ii) the {\em rigid adiabatic 
approximation}. In both cases the eccentricity of the orbit and time dependency of the
arms are ignored. In the long wavelength
approximation the actual size of LISA's arms with respect to $\lambda$ (the wavelength
of the impinging radiation) is also neglected,
implying that the instrument transfer function is assumed to be constant. This is
a very good representation of the LISA output for $f \ll f_{\ast}$. In the
rigid adiabatic approximation the frequency and source position dependence of
the transfer function are fully included; 
the LISA Michelson output that is derived under these assumptions is an excellent 
representation of the exact one up to $f \approx 0.5$ Hz, in the sense that the overlap
between the exact and approximated output is always $\ge 0.97$~\cite{RCP03}. 
In the remainder of the paper we will therefore consider 
the rigid adiabatic approximation as a faithful representation of the signal extracted
at the LISA output, as we are considering binary systems whose radiation is 
at $f < 0.5$ Hz. We refer the reader
to ~\cite{lisa_ppa,Cutler98,CR03,RCP03} and references therein for more details.
Here we briefly review, mainly to establish notation,
the expressions of the signal recorded at the output 
of the LISA Michelson interferometer using these two different approximations.

Consider a generic gravitational wave source whose position in the sky,
with respect to an observer on LISA,
is identified by the unit vector ${\bf \hat N}$. Gravitational waves travel in the
$-{\bf \hat N}$ direction, and are described by the two independent
polarisations $h_{+}$ and $h_{\times}$~\cite{Thorne87}. The metric perturbation $h_{ab}$ at
the detector can be decomposed as the sum of the two independent polarisation states 
according to
\be
h_{ab}(t) = h^{+}(t)\epsilon_{ab}^{+}(t) + h^{\times}(t)
\epsilon_{ab}^{\times}(t)\,,
\quad\quad (a,b = 1,2,3)\,,
\label{hab}
\ee
where $\epsilon_{ab}^{+}$ and $\epsilon_{ab}^{\times}$ are the wave's polarisation
tensors. They can be expressed as a function of the source basis tensors, $e^{+}_{ab}$
and $e^{\times}_{ab}$, and the wave polarisation angle 
$\psi$ as:
\begin{subequations}
\ba
\epsilon^{+}_{ab} & = & \cos2\psi\, e^{+}_{ab} - \sin2\psi\, 
e^{\times}_{ab},
\label{ee+}
\\
\epsilon^{\times}_{ab} & = & \sin2\psi\, e^{+}_{ab} + \cos2\psi\, 
e^{\times}_{ab}\,.
\label{eex}
\ea
\end{subequations}
The source basis tensors are constructed by considering two 
unit vectors $m^a$ and $n^a$ orthogonal to each other
and to the wave propagation direction $-{\bf \hat N}$, in order to form a left-handed
Cartesian tern, according to:
\begin{subequations}
\ba
e^{+}_{ab} & = & m_a m_b - n_a n_b\,,
\label{e+}
\\
e^{\times}_{ab} & = & m_a n_b + n_a m_b\,.
\label{ex}
\ea
\end{subequations}
If the signal is described by the two polarisation amplitudes $A_{+}(t)$ and
$A_{\times}(t)$, respectively, and the gravitational phase 
$\phi_{GW}(t)$, then Eq.~(\ref{hab}) becomes
\be
h_{ab}(t) = \left[A_{+}(t) \epsilon_{ab}^{+}(t) - iA_{\times}(t) 
\epsilon_{ab}^{\times}(t)\right]\,e^{i\phi_{GW}(t)}\,,
\label{hab1}
\ee
where we have explicitly included the dependence on time.

\subsection{The rigid adiabatic approximation}

\begin{figure}
\includegraphics[height=10cm]{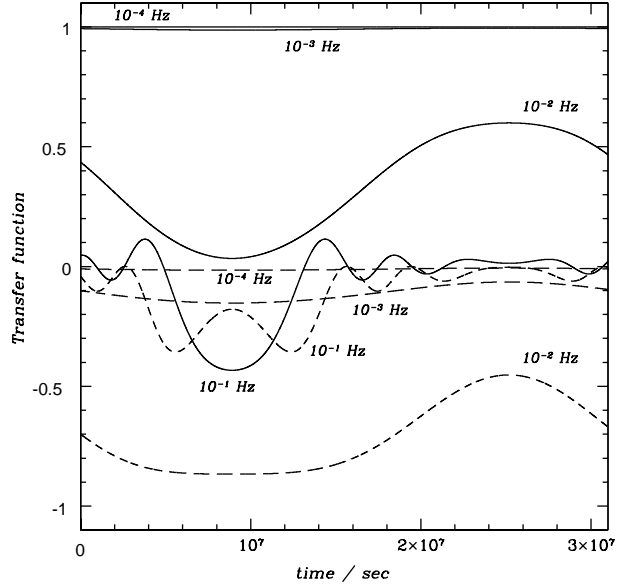}
\caption{
The time evolution of the LISA transfer function for selected frequencies. 
The plot shows the transfer function, cf Eq.~(\protect{\ref{transfer}}), 
as a function of time (for one LISA period,
corresponding to 1 yr) for one of the arms
and a randomly chosen source location. The values of the transfer function for four
selected frequencies (see labels) are shown: $f = 10^{-4}\,{\rm Hz},\ 
10^{-3}\,{\rm Hz},\ 10^{-2}\,{\rm Hz}$ and $10^{-1}\,{\rm Hz}$ 
(solid line: real part; dashed line: imaginary part). Notice that even by 
$f = 10^{-3}$ Hz the 
transfer function is no longer constant over the LISA orbital period.
}\label{fig:transt}
\end{figure}

\begin{figure}
\includegraphics[height=10cm]{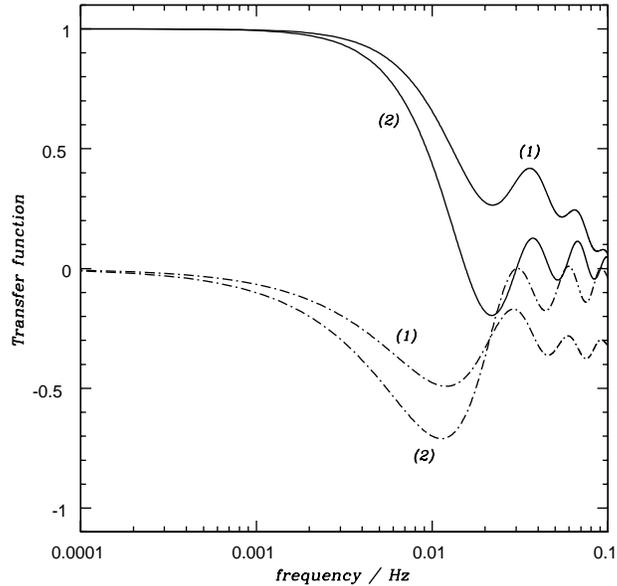}
\caption{
The frequency dependence of the LISA transfer function. The plot shows the
LISA transfer function, cf Eq.~(\protect{\ref{transfer}}), 
as a function of frequency for two (see labels) selected relative
positions and orientations of LISA with respect to a source (solid line: real
part; dotted-dashed line: imaginary part). Notice that at a few mHz the transfer 
function is already significantly different from the constant value apparent at
$f \ll f_{\ast}$.
}\label{fig:transf}
\end{figure}

In the so-called {\em rigid adiabatic approximation}
the strain at the detector induced by the metric perturbation, Eqs.~(\ref{hab}) 
and~(\ref{hab1}), is~\cite{RCP03}
\be
h(t) = D^{ab}(t)\,h_{ab}(t)\,,
\label{h}
\ee
where $ D^{ab}(t)$ is the time dependent detector response tensor. If we consider
the detector formed by the arms $l^a_j$ and $l^a_k$ -- ${\bf \hat l}_{j}\, (j=1,2,3)$ 
are the unit vectors along each of the LISA arms; in a frame attached to the solar 
system barycentre they are time dependent functions as a consequence of LISA's motion -- 
the detector response tensor reads
\be
D^{ab} = \frac{1}{2}\,
(l^{a}_j\, l^{b}_j\,T_{j}-l^{a}_k\, l^{b}_k\,T_{k})\,,
\label{Dab}
\ee
where
\ba
T_{j}  & = & \frac{1}{2}\,
\mathrm{sinc}\left[ \,\frac{f}{2f_{\ast}}\, 
\left(1 + l^{c}_j\,N_c\right)\right]\,
\exp\left\{ 
-i \left[ \,\frac{f}{2f_{\ast}}\left(3- l^c_{j}
N_c\right)\right]
\right\} \nonumber
\\
& & + 
\frac{1}{2}\,
\mathrm{sinc}\left[ \,\frac{f}{2f_{\ast}}\, 
\left(1 - l^{c}_j\,N_c\right)\right]\,
\exp\left\{ 
-i \left[ \,\frac{f}{2f_{\ast}}\left(1-l^{c}_j\,N_c\right)\right]
\right\}\,.
\label{transfer}
\ea
$T_{j}$ is the instrument transfer function, whose behaviour as a function of
time and frequency is shown in Figures~\ref{fig:transt} and~\ref{fig:transf}. 
In general, this is a time dependent function which depends on the source location 
and the instantaneous frequency of the wave; it carries 
key information about the 
interferometer arm length through the characteristic frequency  
$f_{\ast}$, Eq.~(\ref{fast}). It is straightforward to
check, cf Figures~\ref{fig:transt} and~\ref{fig:transf}, that for $f/f_{\ast} \ll 1$, 
$T_{j}$ tends to unity; as the frequency of GWs increases it
develops increasingly stronger oscillations. If we ignore the contribution given
by the geometry of the detector, the transfer function introduces a phase shift $\simgt 1$ rad for $f \simgt 5$ mHz (the LISA peak sensitivity is at $f \simeq 3$ mHz). The overlap of the output~(\ref{h}) 
with the {\em exact} expression of the LISA readout is greater than 0.97 up to 
$\approx 0.5$ Hz~\cite{RCP03}, so that we can 
safely replace the full expression with the rigid adiabatic approximation 
in the frequency range of interest for this paper (0.1 mHz - 0.1 Hz). 

%
%
%
%

%
%
%
%

The expression for the detector response tensor, Eq.~(\ref{Dab}), 
can be recast in the form
\be
D^{ab}(t) = \sum_{n = 1}^{4}\,W^{ab}_{n}(t)\,\,e^{-i\xi_{n}(t)}\,,
\label{D1}
\ee
where
\begin{subequations} 
\ba
W^{ab}_1 & = &  
\frac{1}{4}\,l^{a}_j\, l^{b}_j\,
\mathrm{sinc}\left[ \,\frac{f}{2f_{\ast}}\, 
\left(1 + l^{c}_j\,N_c\right)\right]\,,
\label{W1}
\\
W^{ab}_2 & = &
\frac{1}{4}\,l^{a}_j\, l^{b}_j\,
\mathrm{sinc}\left[ \,\frac{f}{2 f_{\ast}}\, 
\left(1 - l^{c}_j\,N_c\right)\right]\,,
\label{W2}
\\
W^{ab}_3 & = &  
- \frac{1}{4}\,l^{a}_k\, l^{b}_k\,
\mathrm{sinc}\left[ \,\frac{f}{2f_{\ast}}\, 
\left(1 + l^{c}_k\,N_c\right)\right]\,,
\label{W3}
\\
W^{ab}_4 & = &
-\frac{1}{4}\,l^{a}_k\, l^{b}_k\,
\mathrm{sinc}\left[ \,\frac{f}{2f_{\ast}}\, 
\left(1 - l^{c}_k\,N_c\right)\right]\,,
\label{W4}
\ea
\end{subequations}
and
\begin{subequations} 
\ba
\xi_{1} & = & \,\frac{f}{2f_{\ast}}
\left(3 - l_j^{c}\,N_{c}\right)\,,
\label{vp1}
\\
\xi_{2} & = & \,\frac{f}{2f_{\ast}}
\left(1 - l_j^{c}\,N_{c}\right)\,,
\label{vp2}
\\
\xi_{3} & = & \,\frac{f}{2f_{\ast}}
\left(3 - l_k^{c}\,N_{c}\right)\,,
\label{vp3}
\\
\xi_{4} & = & \,\frac{f}{2f_{\ast}}
\left(1 - l_k^{c}\,N_{c}\right)\,.
\label{vp4}
\ea
\end{subequations}
Is is also useful to introduce the {\em generalised antenna
beam patterns} $F_n^{(+)}$ and $F_n^{(\times)}$ defined as
\begin{subequations} 
\ba
F_n^{(+)}(t) & \equiv & W^{ab}_{n}(t)\,
\epsilon^{+}_{ab}(t)\,,
\\
F_n^{(\times)}(t) & \equiv & W^{ab}_{n}(t)\,
\epsilon^{\times}_{ab}(t)\,.
\ea
\end{subequations}
$F_n^{(+)}$ and $F_n^{(\times)}$ 
are time dependent because of the change of orientation of LISA
during the typical observation time $\sim$1 yr.
It is straightforward to verify that at low frequencies
\begin{subequations} 
\ba
F_+ & = & \sum_{n = 1}^4 F_n^{(+)} \quad\quad (f \ll f_{\ast})\,,
\\
F_\times & = & \sum_{n = 1}^4 F_n^{(\times)} \quad\quad (f \ll f_{\ast})\,,
\ea
\end{subequations} 
where $F_+$ and $F_{\times}$ are the usual antenna beam patterns 
cf. Eqs.~(\ref{Fplus}) and~(\ref{Fcross}). 

Using Eqs~(\ref{hab}),~(\ref{D1}), (\ref{W1})-(\ref{W4}) 
and~(\ref{vp1})-(\ref{vp4}), the (real) 
strain at the detector becomes:
\ba
h(t) & = & \sum_{n = 1}^{4} \Bigl(F_n^{(+)}(t)\,A_{+}(t)
\cos[\phi_{GW}(t) - \xi_{n}(t) + \varphi_{\rm D}(t)]
\nonumber
\\
& & \quad + F_{n}^{(\times)}(t)\,A_{\times} (t)
\sin[\phi_{GW}(t) - \xi_{n}(t) + \varphi_{\rm D}(t)]\Bigr)\,,
\label{ha}
\ea
where we have included the phase Doppler shift induced in the signal at the
LISA output by the motion of the
LISA barycentre with respect to a source, see Eq.~(\ref{dopplph}). 
Using double angle formulae one can express Eq.~(\ref{ha})
as the superposition 
of 4 harmonics with different (time dependent) polarisation amplitudes and phases
\be
h(t) = \sum_{n = 1}^4 B_n(t) \cos\chi_n(t)\,,
\label{ha1}
\ee
where
\ba
B_n(t) & = & \left[(F_n^{(+)}(t)\,A_+(t))^{2} + (F_n^{(\times)}(t)\,A_\times(t))^{2}\right]^{1/2}\,,
\label{Bn}
\\
\chi_n(t) & = & \phi_{GW}(t) - \xi_n(t) + \varphi_{\rm D}(t) + \varphi_{n}(t)\,,
\label{chin}
\ea
and
\be
\varphi_n(t) = \arctan\left[-\frac{F_n^{(\times)}(t)\,A_\times(t)}
{F_n^{(+)}(t)\,A_+(t)}\right]\,.
\label{polphasen}
\ee
Explicit expressions for $B_n(t)$, $\chi_n(t)$ and $\varphi_n(t)$ are given in the Appendix.

\subsection{The long wavelength approximation}
\label{subs:lowf}

In the low frequency region of the LISA sensitivity band, {\em i.e.} 
$f/f_{\ast} \ll 1$, the transfer function can be approximated as a 
constant~\cite{RCP03}, cf Figures~\ref{fig:transt} and~\ref{fig:transf}; $T_j$ is therefore independent
of the signal frequency and source position in the sky. The strain at the 
detector output, 
Eq.~(\ref{h}), takes the usual form
\be
h_L(t) = h_+(t) F_+(t) + h_\times(t) F_\times(t)\,.
\label{hl}
\ee
This expression is known as the {\em long wavelength approximation} to the
detector output. In the previous expression $F_+$ and $F_\times$ are the 
antenna beam patterns.
If one takes the observable constructed using the two arms identified by 
${\bf \hat l}_j$ and ${\bf \hat l}_k$, they read
\begin{subequations}
\be
F_{+} =  
\frac{1}{2}(l^{a}_j\, l^{b}_j - l^{a}_k\, l^{b}_k)\epsilon^{+}_{ab} \,,
\label{Fplus}
\ee
\be
F_{\times}  =  
\frac{1}{2}(l^{a}_1\, l^{b}_1 - l^{a}_2\, l^{b}_2)\epsilon^{\times}_{ab}\,. 
\label{Fcross}
\ee
\end{subequations}
As is the case for $F_n^{(+)}$ and $F_n^{(\times)}$ ($n=1,...,4$),
$F_+$ and $F_\times$ are time dependent because of LISA's change in orientation 
with respect to source location during the observation period. 

The detector output~(\ref{hl}) can be cast in the form
\be
h_L(t) = B(t) \cos\chi(t)
\label{hl1}
\ee
where
\ba
B(t) & = & \left[(F_{(+)}(t)\,A_+(t))^{2} + (F_{(\times)}(t)\,A_\times(t))^{2}\right]^{1/2}\,,
\label{B}
\\
\chi(t) & = & \phi_{GW}(t) + \varphi(t) + \varphi_{\rm D}(t)\,,
\label{chi}
\ea
and
\be
\varphi(t) = \arctan\left[-\frac{F_{(\times)}(t)\,A_\times(t)}
{F_{(+)}(t)\,A_+(t)}\right]\,.
\label{polphase}	
\ee
Explicit expressions for $B(t)$, $\chi (t)$ and $\phi (t)$ are given in the Appendix.

\section{Signal detection}
\label{sec:detect}

In this section we investigate the impact of the instrument transfer function
on detection. Due to the essentially perfect knowledge of signal waveforms for
this class of sources,
we assume that the signal processing scheme
will be based on a coherent approach, where the data are correlated
with a discrete bank of templates to extract the signal from the noise. 
The goal of this section is to identify the frequency band over
which the use of the long-wavelength 
approximation $h_L$ for signal templates ({\em i.e.} constant transfer function), 
cf Eq.~(\ref{hl1}),
does not affect the signal-to-noise ratio (SNR) at which the actual radiation 
embedded in the detector noise can be detected. At the LISA output, in fact, gravitational
waves are modulated by the complex structure of the instrument transfer function
and are actually represented by Eq.~(\ref{ha1}). It has been recently suggested~\cite{RCP03}
that the long wavelength approximation, $h_{L}(t)$, is already 
not a good approximation of the exact detector
output at $f\approx 3$ mHz; here we show that for the purpose of detecting
monochromatic signals $h_{L}(t)$ is actually a completely satisfactory approximation 
at least up to $f \approx 10$ mHz, in the
sense that it returns a fitting factor which is always greater than 0.97.
We start by briefly reviewing the key concepts
and formulae of signal detection through matched
filtering -- we refer the reader to~\cite{Helstrom68,WZ62,Cutler98} and
references therein for more details -- 
and then present the results of our analysis.

The signal $s(t)$ registered at the detector output is a superposition of 
noise $n(t)$ and gravitational waves $h(t;{\gras \lambda})$,
\be
s(t) = h(t;{\gras \lambda}) + n(t)\,,
\ee
where ${\gras \lambda}$ represents the vector of the unknown parameters 
that characterise the waveform. In the case of monochromatic sources $h(t)$
is fully described by seven independent parameters: 
signal amplitude $A_{\rm GW}$, frequency $f_0$, arbitrary
initial phase at the beginning the observation $\phi_0$, and four angular parameters 
related to the position of the source in the sky, $\theta_N$ and $\phi_N$, 
and the orientation of the orbital plane, $\theta_L$ and $\phi_L$
(we refer the reader to the Appendix where the explicit dependence of $h(t)$ on
$\gras\lambda$ is presented). The
parameter vector is therefore ${\gras \lambda} = \{A_{\rm GW}, \phi_0, f_0,\theta_N, 
\phi_N,\theta_L, \phi_L\} = \{A_{\rm GW}, \phi_0, f_0,{\gras \theta}\}$, where
${\gras \theta} = \{\theta_N, \phi_N,\theta_L, \phi_L\}$. Our notation reflects the
fact that $A_{\rm GW}$ is not a search parameter, as it determines simply the
signal-to-noise ratio at which detection is made, $f_0$ can be easily searched over
using the Fast Fourier Transform of the data stream and $\phi_0$ is an extrinsic
parameter that can be trivially maximised. 
We assume the noise to be stationary and Gaussian (although we do not expect this
condition to be fully met by the actual data), characterised by
a noise spectral density $S_n(f)$.
In the geometrical approach to signal processing the signal $h(t)$
represents a vector in the signal manifold and the signal parameters 
${\gras \lambda}$ are the coordinates on this manifold. One can introduce
the following inner product between two signals $v$ and $w$~\cite{Cutler98}:
\ba
\left(v|w\right) & = & 2\,\int_{-\infty}^{+\infty} 
\frac{\tilde v^*(f) \tilde w(f)}{S_n(f)}\,df
\nonumber\\
& = & \frac{2}{S_0}\,\int_{-\infty}^{+\infty} 
v^*(t) w(t)\,dt\,,
\label{inner}
\ea
where the second equality follows from Parseval's theorem and $S_0$ is the
(essentially constant) noise spectral density at $f_0$.

According to the definition~(\ref{inner}), the optimal SNR at which $h$ can be detected is
\be
\left(\frac{{\rm S}}{{\rm N}}\right)_{\rm opt}
= \frac{(h|h)}{{\rm rms}[(h|n)]} = (h|h)^{1/2}\,.
\label{snr}
\ee
If $q(t;{\gras \Lambda})$ is the family of templates used to 
search for the class of signals $h(t;{\gras \lambda})$ -- notice that 
${\gras \Lambda}$, the template parameter vector, is not necessarily the same (including its
dimensions)
as ${\gras \lambda}$, and
$q$ does not necessarily belong to the same manifold as $h$ -- the 
adequacy of the template family $q$ to search for the signal family $h$  is given by 
the so-called {\em fitting factor}
(FF), defined as~\cite{Apostolatos96}
\be
{\rm FF}({\gras \lambda}) = 
\max_{{\gras \Lambda}} \left\{\frac{(h({\gras \lambda})|q({\gras \Lambda}))}
{\left[(h({\gras \lambda})|h({\gras \lambda}))\,(q({\gras \Lambda})|q({\gras \Lambda})) \right]^{1/2}}\right\}\,.
\label{ff}
\ee
By definition $0 \le {\rm FF} \le 1$, and the effect of an imperfect matching of
filters with signals translates into a reduction of the maximum SNR at which
a source can be detected according to
\be
\left(\frac{{\rm S}}{{\rm N}}\right) = {\rm FF}\times  
\left(\frac{{\rm S}}{{\rm N}}\right)_{\rm opt}\,.
\label{SNRloss}
\ee
It has become standard
in the gravitational wave data analysis literature to assume that a family of 
filters can be considered adequate if ${\rm FF} \ge 0.97$, which corresponds
to a decrease in detection rate by about 10\%; the detection rate, in fact,
scales as $({\rm FF})^3$. Analogously, the {\em match} M is defined as 
\be
M({\gras \theta},{\gras \Theta}) = \max_{\Delta f_0,\Delta \phi_0} 
\left\{\frac{(h({\gras \theta})|q({\gras \Theta}))}
{\left[(h({\gras \lambda})|h({\gras \lambda}))\,(q({\gras \Lambda})|q({\gras \Lambda})) \right]^{1/2}}
\right\}\,,
\label{match}
\ee
where ${\gras \theta}$ and ${\gras \Theta}$ are the parameter vectors of the signal and template, 
respectively; $\Delta f_0$ and $\Delta \phi_0$ represent the mismatch in extrinsic parameters 
between signal and template.
Notice that here we assume the use of
the power spectrum as the detection statistic, which implies the need to search explicitly
over the four angles 
$\theta_N, \phi_N,\theta_L$ and $\phi_L$. Instead of the power spectrum, one
could actually consider the so-called $F$-statistic~\cite{JKS98,KTV04}, which requires
the maximisation of the detection statistic using a discrete template bank only
over the two angles $\theta_N$ and $\phi_N$. The conclusions
of this section are only marginally affected by our choice of detection statistic.

%
%
\begin{figure}
\begin{center}
\mbox{
\scalebox{0.45}{\rotatebox{360}{\includegraphics{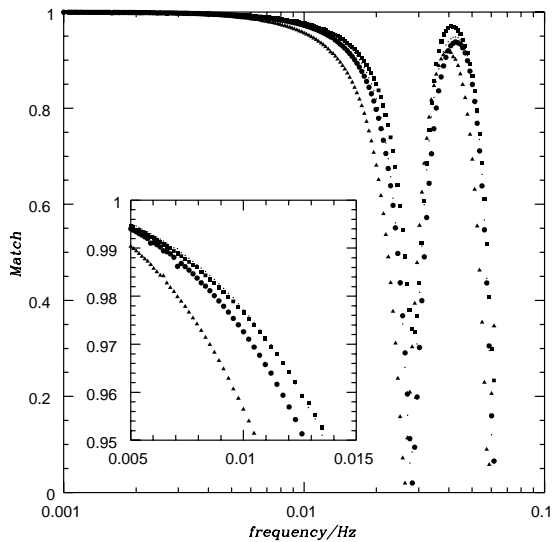}}}}
\end{center}\caption{The frequency dependence of the match. The plot shows the match $M$, 
Eq.~({\protect{\ref{match}}}), as
a function of frequency, for four randomly selected sky positions and orbital
plane orientations 
(dots: $\theta_{N} = 2.5$, $\phi_{N} = 1.4$, $\theta_{L} = 1.2$, and $\phi_{L} = 1.4$;
squares: $\theta_{N} = 1.2$, $\phi_{N} = 4.3$, $\theta_{L} = 0.36$, and $\phi_{L} = 4.2$;
triangles: $\theta_{N} = 3.1$, $\phi_{N} = 0.81$, $\theta_{L} = 2.7$, and $\phi_{L} = 0.79$;
circles: $\theta_{N} = 1.8$, $\phi_{N} = 3.6$, $\theta_{L} = 1.9$, and $\phi_{L} = 3.6$).
The insert offers a zoom of the region around 10 mHz. The observation time is 1 yr.}\label{ffvf}
\end{figure}
%
%

We aim to explore the FF for the family of
filters constructed using the long wavelength approximation. The FF depends on the 
point in parameter space at which it is evaluated and its computation requires
the maximisation over six parameters, cf Eq~(\ref{ff}), which in
turn translates into a substantial computational burden.
In order to investigate extensively the behaviour of templates based
on the long wavelength approximation, while keeping the computational time at a manageable
level, we have actually evaluated, over an extended portion of the whole parameter space,
the match,  Eq.~(\ref{match}), which represents a lower limit
to the FF, and investigated the FF only for selected points. This strategy has the
advantage of reducing substantially the computational burden, while still providing
a reasonable approximation to the FF. 
We therefore set $q = h_{\rm L}$, cf Eq~(\ref{hl1}), 
in Eq.~(\ref{match}) and~(\ref{ff}) and the signal is modelled according to Eq.~(\ref{ha1}). 
All the results presented here assume that
the observation time is $T = 1$ yr. 

\begin{figure}
\begin{center}
\mbox{
\scalebox{0.45}{\rotatebox{360}{\includegraphics{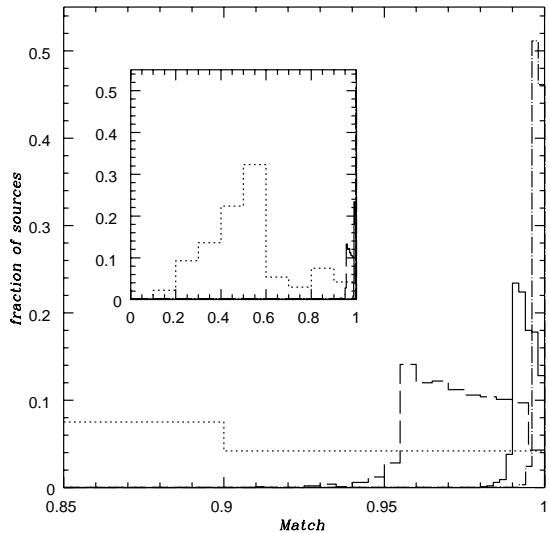}}}}
\end{center}\caption{The distribution of the match. The plot shows the fraction of sources 
for which a given value of the match is attained: in each 
Monte Carlo simulation the match is computed for 1000 sources (all emitting at the same
frequency) with 
random sky position and orientation of the orbital plane. Four values of the gravitational
wave frequency are presented: $f_0 = 3$ mHz (dotted-dashed line), $f_0 = 5$ mHz (solid line),
$f_0 = 10$ mHz (dashed line) and $f_0 = 30$ mHz (dotted line). The observation time
is $T = 1$ yr. The inserted panel shows the same results on a different scale for the
$x$-axis in order to cover the whole range of values of the
match for $f_0 = 30$ mHz.}\label{ffmonte}
\end{figure}

The match and fitting-factor depend on both 
frequency and geometric parameters. In general
we expect the fitting factor to be essentially unity for $f \ll f_{\ast}$, 
but to drop
below 0.97 at some transition frequency whose exact value depends on 
$\theta_{N},\ \phi_{N},\ \theta_{L}$ and $\phi_{L}$. In order to develop some intuition we 
begin by computing the match (we set ${\gras \Theta} = {\gras \theta}$ in Eq.~(\ref{match})) 
for four randomly chosen 
sky positions and orbital plane orientations as a function of frequency; the results 
are plotted in Figure~\ref{ffvf}. 
It is clear that the match falls below 0.97 only at 
about 10 mHz and is not a monotonic function of frequency; in fact there is no maximisation over
the angular parameters and the impact of the transfer function depends on frequency in a non
trivial way. Not surprisingly, the value of the match depends quite significantly on the
position and orientation of a source in the sky. To investigate this dependence we 
select four fiducial frequencies ($f_0 = 3$ mHz, 5 mHz, 10 mHz and 30 mHz) and we
compute the match for 1000 randomly chosen sky positions and wave polarisations at each of
these frequencies~\cite{tsunami}. 
The results are summarised in Figure~\ref{ffmonte} and 
Table~\ref{ffstats}.
It is clear that in the frequency range of peak sensitivity of LISA 
($f\approx 3$ mHz) the
long wavelength approximation is perfectly adequate for signal detection:
{\em all} sources have a match
(and {\em a fortiori} a fitting-factor) larger than 0.99. 
The frequency $f = 10$ mHz seems to mark the transition to the frequency range over 
which the exact LISA response is actually needed in order to recover the full SNR.
Figure~\ref{ffmonte} and Table~\ref{ffstats} show that at 10mHz the match for half
of the sources is below 0.97, but always greater than $\approx 0.91$. In order to check 
whether the fitting-factor, the actual figure of merit we are interested in, shows a 
similar behaviour we have computed FF, Eq.~(\ref{ff}), for the three sources that gave
the three lowest values of the match, 0.914, 0.927 and 0.928. In all three
cases, when the maximisation is carried out over the entire set of parameters, the
fitting-factor is raised above 0.97, and yields 0.976, 0.987 and 0.980, respectively. This
is a very strong indication that FF is indeed $>$ 0.97 at 10 mHz, and one can consider
the long wavelength approximation still adequate at this frequency. 
However, at 30 mHz the picture changes dramatically and essentially all the sources yield
a match which is below 0.97 -- the largest match is 0.976 -- with the lowest value being
as small as 0.074. The mean and median are both $\approx 0.5$. 
In this case too we have computed
the FF for a few sources. Placing the sources in order of ascending match we chose to find
FF for the 1st, 2nd, 300th and 700th entries, which gave matches of 
0.074, 0.109, 0.432 and 0.558 respectively. 
As in the case of $f_0 = 10$ mHz the maximisation over all the parameters increases 
quite substantially
the actual value of the FF, in this case to 0.633, 0.614, 0.673 and 0.737 for the 1st, 2nd,
300th and 700th source, respectively. 
These results indicate that the threshold FF = 0.97, 
which we have set in order to consider the long wavelength approximation
as suitable for detection, is not attainable for the large majority of sources. 
The outcome of our analysis is therefore clear: the long
wavelength approximation, $h = h_L$, Eq.~(\ref{hl1}), can be safely used to construct
{\em detection templates} up to $f_0 =  10$ mHz (for a fitting-factor $> 0.97$), after
which it becomes progressively inaccurate, causing a severe reduction of detection
rate; at $f = 30$ mHz the long wavelength approximation is by far inadequate.

\begin{table} 
\begin{tabular}{|c|cccccc|}
\hline
\hline
$f_0$ & \multicolumn{6}{c|}{Match} \\
(mHz) & Min. & Max. & Median & Mean & Std. Dev. & Fraction$< 0.97$ \\
\hline\hline
3 & 0.993 & 1.000 & 0.998 & 0.998 & 0.001 & 0 \\
\hline
5 & 0.980 & 1.000 & 0.994 & 0.994 & 0.003 & 0 \\
\hline
10 & 0.914 & 0.999 & 0.972 & 0.973 & 0.014 & 0.437 \\
\hline
30 & 0.074 & 0.976 & 0.506 & 0.515 & 0.183 & 0.998 \\
\hline\hline
\end{tabular}
\caption{\label{ffstats} Statistical summary of the values of the match
obtained in each of the Monte Carlo simulations (same as Figure~{\protect{\ref{ffmonte}}})
carried out over 1000 random source positions and orientations for a given gravitational
wave frequency (shown in the first column of the table). The table shows
the minimum (Min.) and maximum (Max.) value of the 
match, cf Eq.~({\protect \ref{match}}), for the different values
of the geometrical parameters $\theta_N$, $\phi_N$, $\theta_L$, $\phi_L$
in each Monte Carlo simulation, and well as the (sample) median, 
mean and standard deviation (Std. Dev.) calculated over the 1000 values
in the simulation. The last column shows the fraction of sources
for which the match is $< 0.97$.}
\end{table}

\section{Parameter estimation}
\label{sec:estimate}

In this section we discuss the errors associated with the parameter 
measurements. The estimation of the minimum mean squared errors with which
source parameters can be extracted in LISA observations of monochromatic
sources has been investigated so far using the long wavelength approximation
across the whole LISA observational window~\cite{Cutler98,CV98,TS02,Seto02MNRAS}. The goal
of this section is to explore to what extent the results reported in previous analyses are affected 
if one actually models the LISA output exactly. Results obtained using the long
wavelength approximation are clearly correct at low frequencies 
($f \ll f_{\ast} \approx 10$ mHz); however 
they become progressively less representative of what one can actually
achieve during the mission in the high frequency portion of
the LISA observational window. Here we establish the frequency
at which the errors significantly differ depending on the approximation and
quantify the discrepancy (as a function of frequency) with respect to previous work.
We start by briefly recalling the general concepts and formulae regarding parameter
estimation -- we refer the reader to~\cite{WZ62,Helstrom68,Cutler98,Finn92,NV98} and 
references therein for more details -- and then present the results of our analysis.

In the limit of large SNR, which applies to the vast majority of signals
considered here, the
errors $\Delta {\gras \lambda}$ associated with the parameters ${\gras \lambda}$ 
that characterise $h(t; {\gras \lambda})$ follow a 
Gaussian probability distribution:
\be
p(\Delta {\gras \lambda}) = \left(\frac{\det({\bf \Gamma})}{2 \,\pi}\right)^{1/2}
\,e^{-\frac{1}{2}\,\Gamma_{jk} \Delta\lambda^j \Delta\lambda^k}\,.
\label{pl}
\ee
In Eq.(\ref{pl}) the matrix $\Gamma_{jk}$ is known as the Fisher information matrix,
which reads~\cite{Cutler98}
\be
\Gamma_{jk}^{(\iota)} \equiv 
\left(\frac{\partial h^{(\iota)}}{\partial \lambda^j} \Biggl|\Biggr.
\frac{\partial h^{(\iota)}}{\partial \lambda^k}\right)\,;
\label{fisher}
\ee
the superscript $\iota = I,II$ labels the detector (as pointed out by 
Cutler in~\cite{Cutler98}, using the three arms
of the LISA instrument one can actually construct two observables, $h^{(I)}$ and $h^{(II)}$,
respectively, that correspond
to two co-located interferometers one rotated by $\pi/4$ with respect to the
other with uncorrelated noise.)
The {\it variance-covariance matrix} is simply given by the inverse of the Fisher
information matrix:
\be
\Sigma^{jk} = \left\m \Delta\lambda^j\,\Delta\lambda^k \right\M = 
\left[\left({\bf \Gamma}^{(\iota)}\right)^{-1}\right]^{jk}\,.
\label{vc}
\ee
The matrix ${\gras \Sigma}$ contains full information about the parameter errors 
and their correlations; in fact the diagonal elements of ${\gras \Sigma}$
represent the expected mean squared errors 
\be
\m (\Delta\lambda^j)^2\M = \Sigma^{jj}\,,
\label{mse}
\ee
and its off-diagonal elements provide information about the correlations
among different parameters through the correlation coefficients $c^{jk}$:
\be
c^{jk} = \frac{\Sigma^{jk}}{\sqrt{\Sigma^{jj}\,\Sigma^{kk}}}\quad\quad (-1 \le c^{jk} \le +1)\,.
\label{corr}
\ee
In the limit of high signal-to-noise ratio, $\Sigma^{jj}$ provides a tight
lower bound to the minimum mean-squared error $\m (\Delta\lambda^j)^2\M$,
the so-called Cramer-Rao bound~\cite{Helstrom68,NV98}. Notice that 
the errors~(\ref{mse}) and the correlation coefficients~(\ref{corr})
depend on the actual value of the signal parameter vector 
${\gras \lambda}$. For the case 
of observations with two or more detectors with uncorrelated
noise, the Fisher information matrix is simply:
$\Gamma_{jk} = \sum_{\iota}\,\Gamma_{jk}^{(\iota)}$. Two parameters
we are interested in computing is the error associated with the position of a
source in the sky (angular resolution) and the orientation of the orbital angular
momentum. Following~\cite{Cutler98} we define these by
\be
\Delta \Omega_{N,L}  = 2 \pi\,
\left\{
\left\m\Delta\cos\theta_{N,L}^2\right\M\,\left\m\Delta\phi_{N,L}^2\right\M -
\left\m \Delta\cos\theta_{N,L}\,\Delta\phi_{N,L} \right\M^2
\right\}^{1/2}\,,
\label{DOmega_N}
\ee
where $N$ labels position and $L$ labels orientation.
The physical meaning of $\Delta \Omega_{N,L}$ is the following: the probability
of ${\bf \hat N}\,,{\bf \hat L}$ to lie {\it outside} an
(appropriately shaped) error ellipse enclosing a solid angle 
$\Delta \Omega$ is  simply $e^{-\Delta \Omega/\Delta \Omega_{N,L}}$. 

Each element of the Fisher information matrix, cf Eqs~(\ref{inner})
and~(\ref{fisher}) can be written as
\be
\Gamma_{jk}^{(\iota)} =
\frac{2}{S_0}\,\int_{-\infty}^{+\infty} 
\partial_j h^{(\iota)}(t)\,
\partial_k h^{(\iota)}(t)\,dt\,,
\label{fishert}
\ee
where $\partial_j \equiv \partial/\partial \lambda^j$. Equivalently, the signal-to-noise ratio is given by 
\be
\left(\frac{{\rm S}}{{\rm N}}\right)^{(\iota)} = 
\frac{2}{S_0}\,\int_{-\infty}^{+\infty} 
[h^{(\iota)}(t)]^2\,
dt\,.
\label{snrt}
\ee
For the problem at hand, for which $f\gg B^{-1}dB/dt$ and $f\gg B_n^{-1}dB_n/dt$,
cf Eq.~(\ref{B}) and~(\ref{Bn}), one can actually simplify the full expression of the
Fisher information matrix reducing it to
\begin{eqnarray}
\Gamma_{jk}  & = & 
\frac{1}{S_{0}}\int_{-\infty}^{\infty}\,
\left[\partial_{j}B(t)\partial_{k}B(t)
+ B^{2}(t)\partial_{j}\chi(t)\partial_{k}\chi(t)\right]\,dt\,,
\label{fishertL}
\end{eqnarray}
in the long-wavelength approximation -- $h = h_L$, cf Eq.(\ref{hl1}) -- and to
\begin{eqnarray}
\Gamma_{jk}  & = & \frac{1}{S_0}\,\sum_{n,m=1}^{4} \int_{-\infty}^{\infty}
\Bigl[\partial_{j}B_{n}(t)\partial_{k}B_{m}
+B_{n}(t)B_{m}(t)\partial_{j}\chi_{n}(t)\partial_{k}\chi_{m}(t)\Bigr]
\cos(\chi_{n}-\chi_{m}) 
\nonumber\\
& & 
\quad\quad\quad +\Bigl[A_{m}(t)\partial_{j}A_{n}(t)\partial_{k}\chi_{m}-
A_{n}(t)\partial_{j}A_{m}(t)\partial_{k}\chi_{n}\Bigr]
\sin(\chi_{n}-\chi_{m})dt\,
\label{fishertA}
\end{eqnarray}
in the rigid adiabatic approximation, where $h$ is given by Eq.~(\ref{ha1}).

\begin{figure}
 \includegraphics[height=10cm]{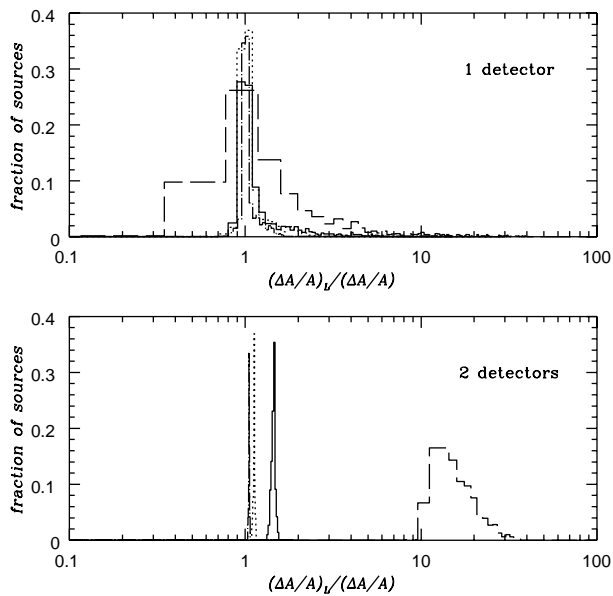}
\caption{\label{fig:histsig} 
Distribution of the ratio between the errors associated with
measurements of the signal amplitude using the long wavelength approximation and the
rigid adiabatic approximation. The histograms show the fraction of 
sources (out of the 1000, with random position and orientation, used in the Monte Carlo 
simulations) for which the amplitude can be measured with a fractional error
characterised by a given value of the 
ratio $(\Delta A/A)_{L}/(\Delta A/A)$, shown on the $x$-axis (see text for more details).  Four
histograms are shown, corresponding to different values of the gravitational
wave frequency $f_0$: 3 mHz (dotted-dashed line), 5 mHz (dotted line), 
10 mHz (solid line) and 30 mHz (dashed line). One year of integration is assumed in
the computation of the Fisher Information matrix. The top and bottom panels refer to measurements
carried out with one and two detectors, respectively. A statistical summary of these results
is presented in Table~{\protect{\ref{tab:ampl}}}.}  
\end{figure}
\begin{table} 
\begin{tabular}{|c|c|cccccc|}
\hline
\hline
$\iota$ & $f$ & \multicolumn{6}{c|}{$(\Delta A/A)_L/(\Delta A/A)$} \\
        & (mHz) & Min. & Max. & Median & Mean & Std. Dev. & Fraction $< 1$ \\
\hline\hline
I    & 3 & 0.756 & 1.642 & 1.007 & 1.049 & 0.124 & 0.40 \\
I+II & 3 & 1.010 & 1.052 & 1.034 & 1.033 & 0.007 & 0.0 \\
\hline
I    & 5 & 0.752 & 2.483 & 1.017 & 1.127 & 0.279 & 0.358 \\
I+II & 5 & 1.055 & 1.127 & 1.096 & 1.094 & 0.012 & 0.0 \\
\hline
I    & 10 & 0.768 & 6.535 & 1.065 & 1.511 & 1.032 & 0.303 \\
I+II & 10 & 1.331 & 1.541 & 1.450 & 1.444 & 0.036 & 0.0 \\
\hline
I    & 30 & 0.392 & 44.164 & 1.599 & 4.949 & 7.696 & 0.219 \\
I+II & 30 & 9.539 & 36.435 & 15.467 & 16.769 & 4.979 & 0.0 \\
\hline
\hline
\end{tabular}
\caption{\label{tab:ampl} Statistical summary of the values of 
$(\Delta A/A)_L/(\Delta A/A)$ obtained in each 
of the Monte Carlo simulations carried out over 
1000 random source positions and orientations for a given
emission frequency (shown in the second column of the table). The table shows
the minimum (Min.) and maximum (Max.) value of 
$(\Delta A/A)_L/(\Delta A/A)$ (see text for more details) for the different values
of the geometrical parameters $\theta_N$, $\phi_N$, $\theta_L$, $\phi_L$
in each Monte Carlo simulation, as well as the (sample) median, 
mean and standard
deviation (Std. Dev.) of $(\Delta A/A)_L/(\Delta A/A)$ calculated over the 1000 values
in the simulation. 
The last column shows the fraction of sources
for which $(\Delta A/A)_L/(\Delta A/A) < 1$, that is the long wavelength
approximation underestimates the errors in the measurement of the
signal amplitude. The values are shown both for the case of observations carried
out with a single interferometer ($\iota = I$) and with a pair of interferometers
($\iota = I + II$). The observation time is one year. Histograms for the
distribution of $(\Delta A/A)_L/(\Delta A/A)$ are presented in Figure~{\protect{\ref{fig:histsig}}}.
}
\end{table}

LISA parameter estimation for monochromatic signals strongly depends 
on the actual value of the 
signal parameters, in particular, emission frequency and 
location and orientation of a source with respect to the detector. This
represents a large parameter space that one needs to
explore in order to obtain meaningful results. We perform this 
exploration by means of Monte-Carlo
simulations~\cite{tsunami}. For each fiducial source we set $f_0$ and randomly select
the geometrical parameters $\theta_N$, $\phi_N$, $\theta_L$, $\phi_L$ 
(and the arbitrary initial phase $\phi_0$) and
compute the inverse of the Fisher information matrix for the cases where $h$ is
given by Eq.~(\ref{hl1}) and Eq.~(\ref{ha1}). The observation time is set to $T = 1$ yr. 
The signals are normalised in such a way that they produce the
same signal-to-noise ratio in both cases. For each parameter 
$\lambda^j$, we can therefore evaluate the ratio
$(\Delta\lambda^j)_L/\Delta\lambda^j$, where the subscript 'L' indicates that the estimated
error is computed using the long wavelength approximation for the LISA output;
$(\Delta\lambda^j)_L/\Delta\lambda^j$ is independent of SNR, and
reflects simply the effect of structure of the transfer function on the
estimation of the signal parameters. If the long wavelength approximation is 
indeed a good approximation for
exploring the quality of LISA astronomy, then $(\Delta\lambda^j)_L/\Delta\lambda^j = 1$ . If
$(\Delta\lambda^j)_L/\Delta\lambda^j > 1 (< 1)$ it means that results that have been 
presented in the literature so far 
overestimate (underestimate) the errors associated with parameter measurements
and one can actually expect better (worse) quality astronomy with LISA.

We have calculated the ratio $(\Delta\lambda^j)_L/\Delta\lambda^j$
for four specific frequencies in the LISA observational window, 
$f_0 = 3\,{\rm mHz}$, $5\,{\rm mHz}$, $10\,{\rm mHz}$
and  $30\,\mathrm{mHz}$ (the same values that we adopted in the study of 
match and fitting factor), for 1000
randomly selected positions and orientations of sources in the sky.
We have also considered both the case for one detector 
and for two detectors at $45^{\circ}$ to each other.

\begin{figure}
 \includegraphics[height=10cm]{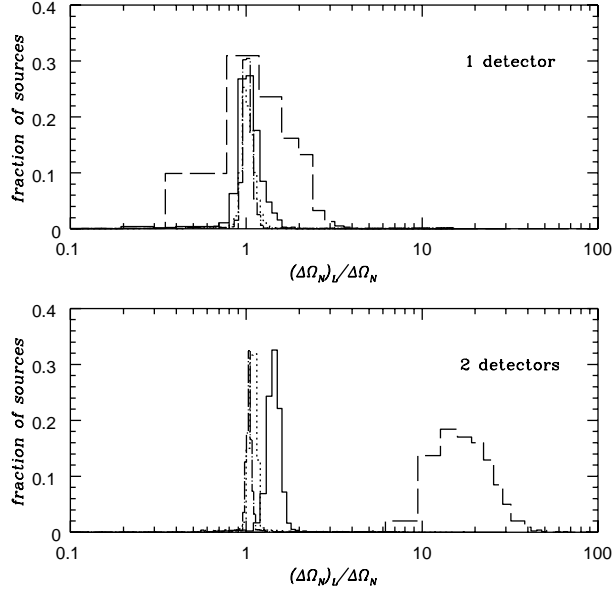}
\caption{\label{fig:histome} 
Distribution of the ratio between the errors associated with
measurements of LISA's angular resolution
using the long wavelength approximation and the
rigid adiabatic approximation. The histograms show the fraction of 
sources that can be resolved in the sky with an angular resolution 
characterised by a given value of the ratio $(\Delta\Omega_{N})_{L}/\Delta\Omega_{N}$,
shown on the $x$-axis (see text for more details). 
The parameters are the same as the ones in Figure~{\protect{\ref{fig:histsig}}}. A statistical
summary of these results is presented in Table~{\protect{\ref{tab:OmegaN}}}.} 
\end{figure}
\begin{table} 
\begin{tabular}{|c|c|cccccc|}
\hline
\hline
$\iota$ & $f$ & \multicolumn{6}{c|}{$(\Delta\Omega_{N})_L/\Delta\Omega_{N}$} \\
        & (mHz) & Min. & Max. & Median & Mean & Std. Dev. & Fraction $< 1$ \\
\hline\hline
I    & 3 & 0.353 & 7.606 & 1.011 & 1.035 & 0.282 & 0.418 \\
I+II & 3 & 0.560 & 10.666 & 1.036 & 1.062 & 0.366 & 0.185 \\
\hline
I    & 5 & 0.292 & 7.737 & 1.019 & 1.047 & 0.290 & 0.403 \\ 
I+II & 5 & 0.561 & 11.832 & 1.099 & 1.127 & 0.408 & 0.380 \\ 
\hline
I    & 10 & 0.208 & 8.144 & 1.045 & 1.089 & 0.331 & 0.360 \\
I+II & 10 & 0.676 & 17.225 & 1.446 & 1.490 & 0.602 & 0.010 \\
\hline
I    & 30 & 0.345 & 14.790 & 1.354 & 1.487 & 0.838 & 0.243 \\ 
I+II & 30 & 6.808 & 259.199 & 18.908 & 20.586 & 12.545 & 0.0 \\
\hline
\hline
\end{tabular}
\caption{\label{tab:OmegaN} 
Statistical summary of the values of 
$(\Delta\Omega_{N})_L/\Delta\Omega_{N}$ obtained in each 
of the Monte Carlo simulations. The parameters of the simulations are
the same as the ones described in Table~\protect{\ref{tab:ampl}}. We refer the reader to
the former table for more details. Histograms for the
distribution of $(\Delta\Omega_{N})_L/\Delta\Omega_{N}$ are presented in 
Figure~{\protect{\ref{fig:histome}}}.}
\end{table}

\begin{figure}
 \includegraphics[height=10cm]{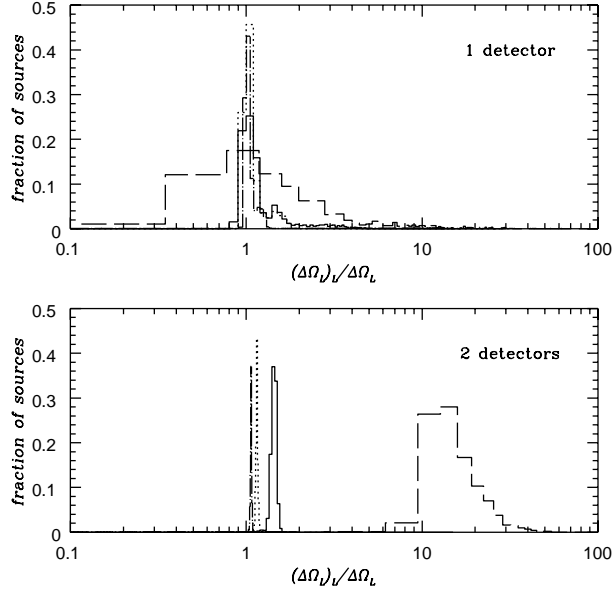}
\caption{\label{fig:histori} 
Distribution of the ratio between the errors associated with
the measurements of the
orientation of the binary orbital plane using the long wavelength approximation and the
rigid adiabatic approximation. The histograms show the fraction of 
sources whose orbital plane orientation, described by the unit vector ${\bf \hat L}$ (cf the Appendix)
can be resolved within a solid angle characterised by a given value of the ratio 
$(\Delta\Omega_{L})_{L}/\Delta\Omega_{L}$, shown on the $x$-axis (cf text for more details). 
The parameters are the same as the ones in Figures~{\protect{\ref{fig:histsig}}}
and~{\protect{\ref{fig:histome}}}. A statistical summary of these results is presented
in Table~{\protect{\ref{tab:OmegaL}}}}    
\end{figure}

We present in detail results for three key parameters:  
amplitude, angular resolution of 
${\bf \hat N}$ and angular resolution of ${\bf \hat L}$ in Figures ~\ref{fig:histsig},
~\ref{fig:histome} 
and ~\ref{fig:histori}, respectively, and Tables~\ref{tab:ampl},
\ref{tab:OmegaN} and~\ref{tab:OmegaL}.
Each plot is a histogram showing the fraction of sources for which
$(\Delta\lambda^j)_L/\Delta\lambda^j$ 
(where $j$ labels the relevant parameter)
falls into each bin (for observations with one and two detectors). 
The tables provide a statistical summary of the outcome of each
Monte Carlo simulation for the three chosen parameters.

The qualitative behaviour of the results is very much consistent with what
one would expect: as the frequency of the signal increases the effect
of the LISA transfer function becomes more pronounced, and the errors on
the source parameters are actually smaller than the ones obtained by
using the simple long wavelength approximation, that is the ratios
$(\Delta A/A)_{L}/(\Delta A/A)$, $(\Delta\Omega_{N})_{L}/\Delta\Omega_{N}$ and
$(\Delta\Omega_{L})_{L}/\Delta\Omega_{L}$ become all greater than 1.
This is clearly due to the structure of the signal recorded at the 
detector which is
amplitude and phase modulated in a characteristic way that depends on
the frequency and source position in the sky. In other words the actual
LISA output provides more discriminating power than the one inferred from
the simple long
wavelength approximation. The values of the mean and median of
$(\Delta A/A)_L/(\Delta A/A)$, $(\Delta\Omega_{N})_L/\Delta\Omega_{N}$
and $(\Delta\Omega_{L})_L/\Delta\Omega_{L}$ are all greater than 1
({\em i.e.} the rigid adiabatic approximation provides errors smaller
than the long wavelength approximation) and increase as the emission
frequency increases (the higher the frequency the stronger the modulations,
cf Figs.~\ref{fig:transt} and~\ref{fig:transf}). Also the fraction of
points for which the above ratios are larger than 1 increases. For $f<
10$ mHz the difference in results between the long wavelength approximation and
the rigid adiabatic approximation is (on average) $\approx 5-15$\%. For $f\geq 10$ mHz
the discrepancy in the errors becomes much more pronounced.
The difference in errors
provided by the two different approximations is particularly strong
when one considers the potential of LISA to behave as 
two interferometers with uncorrelated
outputs: at 10 mHz and 30 mHz respectively the errors are actually 50\% and a factor
$\sim 10$ smaller than the ones obtained using the long wavelength
approximation. However, it is important to stress -- cf the spread in the histograms in 
Figures ~\ref{fig:histsig}, ~\ref{fig:histome} 
and ~\ref{fig:histori}, and the minimum and maximum value of
$(\Delta A/A)_L/(\Delta A/A)$, $(\Delta\Omega_{N})_L/\Delta\Omega_{N}$
and $(\Delta\Omega_{L})_L/\Delta\Omega_{L}$ in Table~\ref{tab:ampl},
\ref{tab:OmegaN} and~\ref{tab:OmegaL} -- that if one considers any
given source, even at $f_0 = 3$ mHz or 5 mHz right in the heart
of the LISA sensitivity band where one would expect the long wavelength
approximation to be completely adequate, the difference in the value
of the errors can range from $\approx 30$\% to a factor $\approx 10$. Moreover,
depending on the source geometrical parameters, the actual errors (the ones
derived using the rigid adiabatic approximation) could be 
{\em larger} than the ones predicted using the long wavelength approximation. 
For detailed studies of LISA astronomy it is therefore important 
to consider the real response of the instrument even in the mHz range.

\begin{table} 
\begin{tabular}{|c|c|cccccc|}
\hline
\hline
$\iota$ & $f$ & \multicolumn{6}{c|}{$(\Delta\Omega_{L})_L/\Delta\Omega_{L}$} \\
        & (mHz) & Min. & Max. & Median & Mean & Std. Dev. & Fraction $< 1$ \\
\hline\hline
I      & 3 & 0.422 & 1.384 & 1.017 & 1.043 & 0.077 & 0.295 \\
I + II & 3 & 0.995 & 1.086 & 1.033 & 1.032 & 0.010 & 0.006 \\
\hline
I      & 5 & 0.915 & 2.312 & 1.036 & 1.110 & 0.192 & 0.260 \\
I + II & 5 & 1.011 & 1.154 & 1.094 & 1.092 & 0.016 & 0.0 \\
\hline
I      & 10 & 0.825 & 5.187 & 1.105 & 1.440 & 0.777 & 0.235 \\
I + II & 10 & 1.062 & 1.614 & 1.442 & 1.435 & 0.053 & 0.0 \\
\hline
I      & 30 & 0.247 & 61.811 & 1.856 & 4.746 & 7.429 & 0.211 \\ 
I + II & 30 & 5.988 & 52.498 & 15.150 & 16.941 & 6.567 & 0.0 \\
\hline
\end{tabular}
\caption{\label{tab:OmegaL} 
Statistical summary of the values of 
$(\Delta\Omega_{L})_L/\Delta\Omega_{L}$ obtained in each 
of the Monte Carlo simulations. The parameters of the simulations are
the same as the ones described in Tables~\protect{\ref{tab:ampl}}
and~\protect{\ref{tab:OmegaN}}. We refer the reader to the former tables for more details.
Histograms for the distribution of $(\Delta\Omega_{L})_L/\Delta\Omega_{L}$ are presented in 
Figure~{\protect{\ref{fig:histori}}}.}
\end{table}

We would like to conclude this discussion by pointing out another
subtlety. So far we have concentrated on the comparison of the lower
bounds on the minimum mean squared errors estimated using two different
expressions of the detector output. The conclusion is that the full 
transfer function needs to be included in order to obtain accurate 
predictions of the quality of LISA astronomy. If one sticks to 
the long wavelength approximation as a model of the detector output,
such predictions become increasingly less accurate as the
frequency of the putative source increases. We have therefore argued that
the errors are (on average) actually smaller than the ones reported so far in the 
literature. However, the actual errors with
which LISA will be able measure source parameters do not necessarily decrease
at higher frequencies: in fact, for
a given distance, position and orientation in the sky of a monochromatic
source, when the frequency increases also the LISA noise spectral density
becomes larger (which in turn degrades the quality of measurements).
There is therefore a competition of effects, and it is
simple to show that eventually the signal-to-noise degradation 
``wins'' over richness of signal structure.

\section{Conclusions}
\label{sec:concl}

We have explored the effect of the LISA transfer function on signal detection
and parameter estimation for monochromatic sources. We have shown that for 
signal detection, although the long wavelength approximation is not a faithful
representation of the Michelson interferometer output for $f\simgt 3$ mHz,
it can actually be safely used in constructing templates for monochromatic signals
up to 10 mHz. The fitting-factor, in fact, always exceeds 0.97 in this 
frequency regime. For parameter estimation, on the other hand, the effects of
the frequency and position dependent transfer function already become significant 
around 3-to-5 mHz and the use of the long wavelength approximation is totally 
inappropriate for $f\simgt 10$ mHz: the estimate of the measurement errors
can be quite different, ranging from 5\% to a factor $\sim 10$ (or more), from
the ones obtained using the low frequency approximation depending on the source parameters
and the observational mode (LISA as a 2 or 3 arm instrument).
Our analysis is limited in two main respects. In the study of signal detection
we have explored the parameter space
quite extensively for the calculation of the match, but not for the fitting factor; this
is entirely due to the computational constraints, but can be alleviated
in the future by accommodating longer runs. Secondly, both for parameter estimation 
and signal detection, we have restricted attention only to monochromatic signals. 
If the source chirp mass is, say,
$1\,\Ms$ and the observation time 1 yr, the radiation becomes linearly chirping at $f \approx 5$ mHz.
This introduces an additional  parameter (the first time derivative of the frequency) in
the signal gravitational wave phase. The transition to a linearly chirping regime  depends on 
the source chirp mass and is at higher frequencies for white dwarf binary systems. Therefore,
depending on the source mass that is considered, the analysis presented in this paper might require 
the addition of one parameter in the signal waveform. However, it is clear that the results 
presented here would not be affected in any significant way, and the general conclusions regarding
the effect of the transfer function would apply. We plan to return to these issues in a future paper.

\begin{acknowledgements}

We would like to thank C. Cutler for several discussions about LISA observations. We are
also grateful to A. Mercer, C. Messenger and R. Vallance for assisting us in the
early stages of implementation of the Monte Carlo simulations on Tsunami.

\end{acknowledgements}

\appendix
\section{Formulae}

Here we present explicit analytical expressions for the rigid adiabatic approximation $h(t)$,
see Eqs.~(\ref{ha1})-(\ref{polphasen}), and the long wavelength approximation $h_L(t)$,
see Eqs.~(\ref{hl1})-(\ref{polphase}), to the LISA detector output 
that are discussed in Section~\ref{sec:output}.

It is convenient to introduce two Cartesian reference
frames (cf, {\em e.g.},~\cite{Cutler98}): a "barycentric" frame 
$(x,y,z)$ tied to the Ecliptic and centred
in the Solar System Barycentre, with ${\bf \hat z}$ perpendicular to 
the Ecliptic, and the plane $(x,y)$ in the Ecliptic itself; a
detector reference frame $(x',y',z')$, centred in the LISA centre of mass and 
attached to the detector, with ${\bf \hat z}'$ perpendicular to the plane
defined by the three arms and the $x'$ and $y'$ axis defined so that 
the unit vectors ${\bf \hat l}_j$ ($j = 1,2,3$) along each arm read
\ba
{\bf \hat l}_j & = & \cos\left[\frac{\pi}{12} + \frac{\pi}{3}\,(j-1)\right] 
{\bf \hat x'}
+ \sin\left[\frac{\pi}{12} + \frac{\pi}{3}\,(j-1)\right] 
{\bf \hat y'}
\nonumber\\
& =  &
\left[\frac{1}{2}\,\sin\alpha_j(t)\,\cos\Phi(t) - \cos\alpha_j(t)\,\sin\Phi(t)\right]
\,{\bf \hat x} \nonumber\\
& & +
\left[\frac{1}{2}\,\sin\alpha_j(t)\,\sin\Phi(t) + \cos\alpha_j(t)\,\cos\Phi(t)\right]
\,{\bf \hat y} 
+ \left[\frac{\sqrt{3}}{2}\,\sin\alpha_j(t)\right]\,{\bf \hat z} \,,
\label{larm}
\ea
where $\alpha_j(t)$ increases linearly with time, according to
\be
\alpha_j(t) = n_{\oplus} t - (j-1)\pi/3 + \alpha_0 \,.
\label{alphaj}
\ee
In the previous expression $\alpha_0$ is a constant specifying the orientation of
the arms at the arbitrary reference time $t=0$, and 
$n_{\oplus} \equiv 2\pi/1\,{\rm yr}$. 
In the Ecliptic frame the motion of LISA's centre-of-mass is described by the 
polar angles 
\ba
\Theta  & = & \frac{\pi}{2}\,,\nonumber\\
\Phi(t) & = & \Phi_0 + n_{\oplus} t\,,
\label{barlisa}
\ea
and the normal to the detector plane ${\bf \hat z'}$ precesses around
${\bf \hat z}$ according to
\be
{\bf \hat z'} = \frac{1}{2} {\bf \hat z} -\frac{\sqrt{3}}{2}
\left[\cos\Phi(t) {\bf \hat x} + \sin\Phi(t) {\bf \hat y}\right]\,.
\label{zprimo}
\ee
We shall follow the convention that primed and unprimed quantities refer to
the frame attached to LISA and the Solar System Barycentre, respectively. 
The geometry of a binary system with respect to LISA is described by the two unit vectors 
${\bf \hat N}$ and ${\bf \hat L}$, where the former 
identifies the source position in the sky and the latter defines the 
direction of the orbital angular momentum (spins are negligible for sub-solar mass 
binary systems and ${\bf \hat L}$ can be therefore regarded as constant). With respect
to an observer on LISA, i.e. the reference frame $(x',y',z')$, the polar coordinates 
of ${\bf \hat N}$ and ${\bf \hat L}$ are $(\theta_{N}',\phi_{N}')$ and $(\theta_{L}',\phi_{L}')$,
respectively. We also define $\psi'$ as 
the time dependent polarisation angle. 
The angles  
$\theta_N'$, $\phi_N'$ and $\psi'$ can be written as a function of the angles measured 
with respect to the solar system barycentre as
\ba
\cos\theta_N'(t) & = & \frac{1}{2} \cos\theta_N - \frac{\sqrt{3}}{2}
\sin\theta_N \cos(\Phi(t) - \phi_N) \,,
\\
\label{thetaNl'}
\phi_N'(t) & = & \Xi_1 + \frac{\pi}{12}
+ \tan^{-1}
\left\{\frac{\sqrt{3}\cos\theta_N +
\sin\theta_N \sin(\Phi(t) - \phi_N)}
{2\sin\theta_N \sin(\Phi(t) - \phi_N)}\right\}\,,\\
\label{phiNl'}
{\rm tan}\,\psi_N' & = & \frac{{\bf \hat L} \cdot {\bf \hat z}' - 
({\bf \hat L} \cdot {\bf \hat N})\,({\bf \hat z}' \cdot {\bf \hat N})}
{{\bf \hat N} \cdot ({\bf \hat L} \times {\bf \hat z}')} \,,
\label{psiNl'}
\ea
where
\be
\Xi_j = n_{\oplus} t - \frac{\pi}{12} - \frac{\pi}{3} (j-1) +\Xi_0\,,
\label{Xijl'}
\ee
and $\Xi_0$ sets the orientation of ${\bf \hat l}_j$ at $t = 0$. 

The relevant scalar and vector products entering the definition of $\psi'$
are 
\ba
{\bf \hat L} \cdot {\bf \hat z}' & = & \frac{1}{2} \cos\theta_L - \frac{\sqrt{3}}{2}
\sin\theta_L \cos(\Phi(t) - \phi_L)\,,
\label{thetaL'}
\\
{\bf \hat L} \cdot {\bf \hat N} & = &
\cos\theta_L \cos\theta_N 
+ \sin\theta_L\sin\theta_N \cos(\phi_L-\phi_N)\,,
\label{LdotNpr}
\\
{\bf \hat{N}} \cdot \left({\bf \hat{L}} \times {\bf \hat{z}}'\right) 
& = &
\frac{1}{2}\sin\theta_N \sin\theta_L \sin ( \phi_L - \phi_N)
\nonumber\\
& & -\frac{\sqrt{3}}{2} \cos\Phi(t) \Bigl(\cos\theta_L \sin\theta_N \sin\phi_N 
- \cos\theta_N \sin\theta_L \sin\phi_L\Bigr)
\nonumber\\
& & - \frac{\sqrt{3}}{2}\sin\Phi(t)\Bigl(
\cos\theta_N \sin\theta_L \cos\phi_L 
- \cos\theta_L \sin\theta_N \cos\phi_N\Bigr)\,.
\label{NdotLvectz'pr}
\ea

The scalar products ${\bf \hat l_j} \cdot {\bf \hat N}\,(j = 1,2,3)$ entering the expression of the
LISA transfer function, Eqs~(\ref{transfer}), (\ref{W1})-(\ref{W4})
and (\ref{vp1})-(\ref{vp4}) can be computed in a straightforward way
using Eqs.~(\ref{larm}), (\ref{thetaNl'}) and~(\ref{phiNl'}). 

The wave polarisation amplitudes, that enter the expressions~(\ref{Bn}), (\ref{polphasen}), (\ref{B})
and (\ref{polphase}),  are given 
by
\begin{subequations}
\ba
A_+(t) & = & 2\, \frac{\Mc^{5/3}}{D}\,
\left[ 1 + \left({\bf \hat L} \cdot {\bf \hat N}\right)^2\right]\,
\left(\pi\,f\right)^{2/3}\,,
\label{Aplus}
\\
A_{\times}(t) & = & - 4\, \frac{\Mc^{5/3}}{D}\,
\left({\bf \hat L} \cdot {\bf \hat N}\right)\,
\left(\pi\,f\right)^{2/3}\,,
\label{hcross}
\ea
\end{subequations}
where $\Mc$ is the source chirp mass, defined as $\Mc = m^{2/5}\,\mu^{3/5}$ --
$m$ and $\mu$ are the total and reduced mass of a binary systems, respectively --,
$f$ is the GW frequency and $D$ is the (luminosity) distance. The GW phase in 
Eqs.~(\ref{chin}) and~(\ref{chi}) is $\phi_{\rm GW}(t) = 2\pi f_0 t - \phi_0$. 

The Doppler phase modulation, in Eqs.~(\ref{ha}), (\ref{chin}) and~(\ref{chi}), is 
\be
\varphi_{\rm D}(t) = 2 \pi R_{\oplus}\, f\,\sin \theta_N \cos(\Phi (t)-\phi_N)\,,
\label{dopplph}
\ee
where $R_{\oplus} = 1\,{\rm AU}$.

{}

\end{document}